\documentclass[prc,superscriptaddress]{revtex4}
\usepackage{mathtools}
\usepackage{graphicx}
\usepackage{listings}
\usepackage{color,soul}
\begin{document}

\title{Exploring nonnormality in magnetohydrodynamic rotating shear flows: application to astrophysical accretion disks}
\author{Tanayveer Singh Bhatia$^{*\dagger}$, Banibrata Mukhopadhyay$^{\dagger}$\\
$^*$Birla Institute of Technology \& Science, Pilani, Rajasthan 333031, India, $^{\dagger}$Department of Physics, Indian Institute of Science, Bangalore 560012, India\\
tanayveer1@gmail.com, bm@physics.iisc.ernet.in}

\begin{abstract}
\noindent
The emergence of turbulence in shear flows is a well-investigated field. 
Yet, there are some lingering issues that have not been sufficiently resolved. 
One of them is the apparent contradiction between the results of linear stability 
analysis quoting a flow to be stable and experiments and simulations proving it to be 
otherwise. There is some success, in particular in astrophysical systems,
based on Magneto-Rotational Instability (MRI), revealing turbulence. However, MRI 
requires the system to be weakly magnetized. Such instability is neither 
a feature of general magnetohydrodynamic (MHD) flows nor revealed in purely hydrodynamic flows. 
Nevertheless, linear perturbations of such flows are nonnormal in nature which 
argues for a possible origin of nonlinearity therein. The concept behind this is that
nonnormal perturbations could produce huge transient energy growth (TEG), which may lead
to non-linearity and further turbulence. However, so far, nonnormal effects in shear flows have 
not been explored much in the presence of magnetic fields. In this spirit, here we 
consider the perturbed visco-resistive MHD shear flows with rotation in general. 
Basically we recast the magnetized momentum balance and associated equations 
into the magnetized version of Orr-Sommerfeld and Squire equations and their magnetic analogues. 
We also assume the flow to be incompressible and in the presence of Coriolis effect solve the 
equations using a pseudospectral eigenvalue approach. We investigate the possible emergence of instability and 
large TEG in three different types of flows, namely, the Keplerian flow, the 
Taylor-Couette (or constant angular momentum) flow and plane Couette flow. We show that, 
above a certain value of magnetic field, instability and TEG both stop occurring. We also 
show that TEG is maximum in the vicinity of regions of instability in the wave number space 
for a given magnetic field and Reynolds number, leading to nonlinearity and plausible
turbulence. Rotating shear flows are ubiquitous in astrophysics, especially accretion disks, 
where molecular viscosity is too low to account for observed data. The primary accepted 
cause of energy-momentum transport therein is turbulent viscosity. Hence, these results would have important 
implications in astrophysics. 
\end{abstract}


\maketitle

\section{Introduction}


The origin of linear instability and turbulence, and subsequent angular momentum transport in various 
classes of shear flows, specifically in astrophysical accretion disks, which are rotating 
shear flows, has not been explained completely yet. However, it is understood from observed data that, 
to explain the accretion in astrophysical disks, some sort of viscosity is required. In the absence of 
adequate molecular viscosity \cite{pringle1981}, turbulent viscosity was argued to play the main role 
in the accretion process by Shakura and Sunyaev \cite{shasun1973}. Nevertheless, a Keplerian accretion 
disk is linearly stable, thus proving it difficult to explain the origin of turbulence in the absence of any 
unstable linear perturbation. Similar problem exists in some laboratory flows. For example, 
plane Poiseuille flow becomes turbulent in the laboratory at a Reynolds number $Re\sim 1000$,
whereas linear theory predicts it to be stable up to $Re=5772$.
An even more severe discrepancy, which has a direct interest to astrophysics, occurs in the case 
of plane Couette flow, which is shown to be turbulent for $Re$ as small as $350$ in laboratory
experiments and numerical simulations. However theoretical analysis shows it to be linearly 
stable for all $Re$ up to infinity. 
Subsequently, Balbus and Hawley applied the idea of Magneto-Rotational 
Instability (MRI) \cite{balbushawley1991}, established originally by Velikhov \cite{velikhov1959} and Chandrasekhar \cite{chandrasekhar1960},
 to resolve the issue of instability and turbulence in magnetized flows and, hence, in some kinds of
accretion disks. But the puzzle remains in laboratory flows which are colder and, hence, MRI
would not work there. Moreover, to work MRI successfully, magnetic field strength has to be weak
(weaker than a critical value depending on $Re$ \cite{nath2015}). Hence, for global purposes, a full
scale exploration of magnetohydrodynamic (MHD) flows is needed. 

Exploration of MHD instabilities in various fluid systems is nothing new. The comprehensive
descriptions of their various properties including eigenspectra of perturbation and stability are 
given in \cite{mhdbook1,fusion} in the limit of ideal MHD and in \cite{mhdbook2} in the 
presence of visco-resistive effects. 
The properties of eigenspectra and instability have also been explored 
to a great degree, even in two dimensions, in the context of tokamak fusion 
physics (see, e.g., \cite{mhdbook2}). Moreover, ideal MHD spectra for cylindrical plasma column 
were explored in order to investigate that how the local criteria govern the existence of the 
accumulating eigenmodes \cite{jpp,pp}. In the context of astrophysical accretion disks
and other transonic flows, full scale MHD instability was found in radially stratified flows
\cite{apjl} as well as in axisymmetric plasmas having poloidal flow speed exceeding critical
slow magnetosonic speed \cite{ppastro}. In a completely different approach, 
MHD instability and plausible turbulence were also argued in accretion disks and other magnetized flows
by computing various types of correlation of perturbations \cite{sujitamit1,sujitamit2}.

Generally, below a certain critical value of $Re$ ($Re_c$), the linear stability analysis 
would predict a flow to be stable, but sometimes the most minutely controlled experiments would result in 
turbulence below $Re_c$ set by the theory. That exactly is being observed in laboratory experiments 
and numerical simulations of plane Poiseuille flow mentioned above, when its 
$Re_c=5772$ \cite{orszag1971}. Such a discrepancy would lead one to believe that simple linear stability 
analysis is probably not the best tool to enlighten the onset of turbulence. In a related field, 
Trefethen, Embree, Schmid and Henningson \cite{trefethenembree,schmidhenningson,
orszag1972} explored the idea 
of nonnormality.
Under this idea, it 
is shown that even in the complete absence of a linearly unstable mode, perturbations could exhibit 
`Transient Energy Growth (TEG)' \cite{reddyhenningson1993}. This happens when the eigenfunctions of a 
linear system are not completely orthogonal in nature and, because of that, certain combinations of the 
eigenfunctions and initial conditions, may develop a significant 
amplitude of (transient) energy growth, despite being stable overall. This form of growth, as the name suggests, occurs only for a 
short period of time, but its magnitude could be sufficient (depending upon the parameters of flows) 
to cause nonlinearity and plausible turbulence in fluid flows.

In this work, we consider the visco-resistive (including fluid viscosity and magnetic diffusivity) 
MHD equations for three cases of flows: with and without the presence of Coriolis (rotational) effects, 
to explore their linear stability and TEG analyses. We precisely consider a small section of
\begin{itemize}
\item plane Couette flow,
\item Keplerian flow,
\item constant angular momentum flow or classic Taylor-Couette flow.
\end{itemize}
The second class of flow often mimics a small section of an astrophysical accretion disk and, hence, 
our results, as will be shown, have important implications in astrophysics. The present work 
is the sequel of the work \cite{nath2015} by the present group towards the application of nonnormality to MHD
shear flows, including astrophysical flows. In the latest work, the authors approached the problem 
in the Lagrangian formulation. While that is an elegant way of approaching it, in particular 
for the purposes of that work, to uncover certain other physics, Eulerian approach is more useful.
Hence, in the present work, we undertake the Eulerian approach to fulfill the underlying physics. 
Overall, the latest work \cite{nath2015} and the present one complement to each other, to 
understand the full picture of the problem.

We begin with the description of model with basic equations 
in \S \ref{basiceq}, followed by perturbed fluid equations in \S \ref{perturbeq}. We then describe these 
equations in an eigenvalue formulation in \S \ref{eigform}, introduce and apply the concept of TEG 
to them in \S \ref{transen} and discuss the numerical considerations used to solve the problem 
(for eigenvalue formulation and TEG) in \S \ref{numcond}. Subsequently, we explore a simpler analytical 
scheme in \S \ref{anal}, which is 
useful to interpret and understand the numerical results presented 
in \S \ref{numrel}. Finally we end with a summary and conclusion in \S \ref{summa}. 

\section{Model}

\subsection{Basic Equations} \label{basiceq}

We consider a small section of a shear flow (shearing box) including rotational effect at a distance $r_0$ from the 
center of flow (e.g. compact object/star in the cases of accretion disks) of size $L$ in the $r$-direction.
The background unperturbed velocity (in the limit $L \ll r_0$ corresponding to linear shear) and magnetic fields 
are respectively given as
\begin{equation} \label{vback}
\overrightarrow{V}=\left({0,-\frac{U_0 X}{L},0}\right),
\end{equation}
\begin{equation} \label{bback}
\overrightarrow{B_0}=({B_1,B_2,B_3}),
\end{equation}
which are generally the solutions of unperturbed momentum balance equations,
where $U_0$ is the background flow speed at the boundaries of shearing box, describing by the local
Cartesian coordinates, in the $r$-direction (locally $X$-direction).
Now the Navier-Stokes equations 
with magnetic body force in the rotating frame of reference, the induction equation, the continuity and solenoidal 
conditions (in CGS units, unless otherwise stated), for the unperturbed flow in the shearing box are given by
\begin{equation} \label{ns}
	{\left(\frac{\partial}{\partial T} + \overrightarrow{V} . \overrightarrow{\nabla'}\right) \overrightarrow{V} + (2 \overrightarrow{\omega} \times \overrightarrow{V})}
	{+ \nabla \left(\frac{P}{\rho}\right) - \frac{1}{4\pi \rho}(\overrightarrow{\nabla'}\times \overrightarrow{B_0})\times \overrightarrow{B_0}=\nu \nabla'^2 \overrightarrow{V}},
\end{equation}
\begin{equation} \label{ind}
	{\frac{\partial \overrightarrow{B_0}}{\partial T} + \overrightarrow{\nabla'} \times (\overrightarrow{B_0} \times \overrightarrow{V}) = \eta \nabla'^2 \overrightarrow{B_0}},
\end{equation}
\begin{equation} \label{cont}
	\overrightarrow{\nabla'}.\overrightarrow{V}=0,
\end{equation}
\begin{equation} \label{sol}
	\overrightarrow{\nabla'}.\overrightarrow{B_0}=0,
\end{equation}
where $\vec\omega$ is the Coriolis vector, arisen due to angular velocity of the fluids, defined as
\begin{equation}
	{\vec\omega}=(0,0,\Omega_0), \quad \Omega_0=\frac{U_0}{qL}, \quad \Omega (r)=\Omega_0 {\left(\frac{r_0}{r}\right)^q}, \quad r_0-r=L,
\end{equation}
$P$ is the total fluid pressure due to all external body forces
including that due to central gravity, $\rho$ is the fluid density, $T$ is time and 
$\nabla' = (\partial/\partial X,\partial/\partial Y,\partial/\partial Z)$, $\nu$ is the kinematic viscosity, $\eta$ is the magnetic diffusivity and $q$ parameterizes the shearing in the flow, with $q=3/2$ corresponding to Keplerian Disk, $q=2$ corresponding to constant angular momentum or Taylor-Couette flow and $q\to \infty$ corresponding to plane Couette flow.
Note that, for a constant $\overrightarrow{B_0}$ to satisfy eq. (\ref{ind}), $B_1$ has to be $0$. For other
details, see \cite{man2005,nath2015}.

For the sake of convenience, we recast the equations in the dimensionless form such that
\begin{equation} \label{nodim}
	X=xL, \, Y=yL, \, Z=zL, \, \overrightarrow{V}=\overrightarrow{U} U_0, \, T=tL/U_0, \,  \overrightarrow{B_0}=\overrightarrow{B_p}\sqrt{\rho}U_0,
\end{equation}
which immediately leads to $\overrightarrow{U}=(0,-x,0)$.

\subsection{Perturbation Equations}\label{perturbeq}

Now, following previous work \cite{mc2013,nath2015}, the perturbed fields, when velocity perturbation $\vec{v}=(u,v,w)$
and magnetic field perturbation $\vec{B}=(B_x,B_y,B_z)$, can be substituted in eqs. 
(\ref{ns}), (\ref{ind}), (\ref{cont}) and (\ref{sol}) and further linearizing them for a constant
background magnetic field to give rise to the following perturbation 
equations in dimensionless forms as 
\begin{equation} \label{nodimns}
\frac{\partial \overrightarrow{v}}{\partial t} + \overrightarrow{U}.\nabla \overrightarrow{v} 
+\frac{2 \hat k \times \overrightarrow{v}}{q} + (\nabla p_{tot}) -  
\frac{1}{4 \pi}(\overrightarrow{B_p}.\overrightarrow{\nabla})\overrightarrow{B} 
= \frac{1}{Re}\nabla^2\overrightarrow{v},
\end{equation}
\begin{equation} \label{nodimind}
\frac{\partial \overrightarrow{B}}{\partial t} + (\overrightarrow{U}.\overrightarrow{\nabla})\overrightarrow{B} 
- (\overrightarrow{B_p}.\overrightarrow{\nabla})\overrightarrow{v}
=\frac{1}{Rm}\nabla^2\overrightarrow{B},
\end{equation}
\begin{equation}
\nabla.\vec{v}=0,
\end{equation}
\begin{equation}
\nabla.\vec{B}=0,
\end{equation}
where $\hat{k}$ is unit vector in $z$-direction and $p_{tot}$ is the total pressure including the 
magnetic contribution in units of $\rho$, $Re=U_0L/\nu$, magnetic Reynolds number
$Rm=U_0L/\eta$ and $\nabla=(\partial/\partial x,\partial/\partial y,\partial/\partial z)$. 
On expanding eq. (\ref{nodimns}) in the components of $x$-, $y$- and $z$-directions, differentiating each of the 
them, respectively, with respect to $x$, $y$ and $z$ and adding them up, we obtain
\begin{equation}
\nabla^2 p_{tot}=\frac{2}{q}\left(\frac{\partial v}{\partial x}-\frac{\partial u}{\partial y}\right) + \frac{\partial u}{\partial y}.
\label{psq}
\end{equation}
Now, taking the Laplacian on both sides of $x$-component of eq. (\ref{nodimns}) and using eq. (\ref{psq}), we obtain
\begin{equation}
\left(\frac{\partial}{\partial t} - x{\frac{\partial}{\partial y}}\right)
\nabla^2 u + \frac{2}{q}\left(\frac{\partial\zeta}{\partial z}\right) - \frac{1}{4 \pi}(\overrightarrow{B_p}.\overrightarrow{\nabla}) \nabla^2 B_x = \frac{1}{Re}\nabla^4 u,
\label{os}
\end{equation}
where $\zeta$ is the $x$-component of vorticity. Further,
differentiating the $z$-component of eq (\ref{nodimns}) with respect to $y$ and $y$-component with respect to $z$ and 
subtracting the two equations directly gives 
\begin{equation} \label{sq}
	\left(\frac{\partial}{\partial t} - x \frac{\partial}{\partial y}\right) \zeta + \left(1-\frac{2}{q}\right) \left({\frac{\partial u}{\partial z}}\right) - \frac{1}{4 \pi}(\overrightarrow{B_p}.\overrightarrow{\nabla}) \zeta_B = \frac{1}{Re}\nabla^2 \zeta,
\end{equation}
where $\zeta_B$ is the $x$-component of magnetic vorticity. Finally, the $x$-component of eq. (\ref{nodimind}) 
gives 
\begin{equation} \label{mos}
	\left(\frac{\partial}{\partial t} - x \frac{\partial}{\partial y}\right)B_x - (\overrightarrow{B_p}.\overrightarrow{\nabla}) u = \frac{1}{Rm}\nabla^2 B_x,
\end{equation}
and following the similar procedure, as followed to obtain eq. (\ref{sq}), for $y$ and $z$-components of eq. 
(\ref{nodimind}) gives 
\begin{equation} \label{msq}
	\left(\frac{\partial}{\partial t} - x \frac{\partial}{\partial y}\right) \zeta_B + \left({\frac{\partial B_x}{\partial z}}\right) - (\overrightarrow{B_p}.\overrightarrow{\nabla}) \zeta = \frac{1}{Rm}\nabla^2 \zeta_B,
\end{equation}
where eqs. (\ref{os}) and (\ref{sq}) resemble the Orr-Sommerfeld and Squire equations, respectively, along with 
the contributions from magnetic field and rotation, and eqs. (\ref{mos}) and (\ref{msq}) represent their magnetic 
analogues.

The boundary conditions, because of the no-slip assumption, are
\begin{equation} \label{bc}
	u=\frac{\partial u}{\partial x}=B_x=\frac{\partial B_x}{\partial x}=\zeta=\zeta_B=0 \quad {\rm at} \, x=\pm 1,
\end{equation}

\subsection{Eigenvalue Formulation} \label{eigform}

\begin{figure}[h]
	\centering
	\fbox{\includegraphics[width=\linewidth]{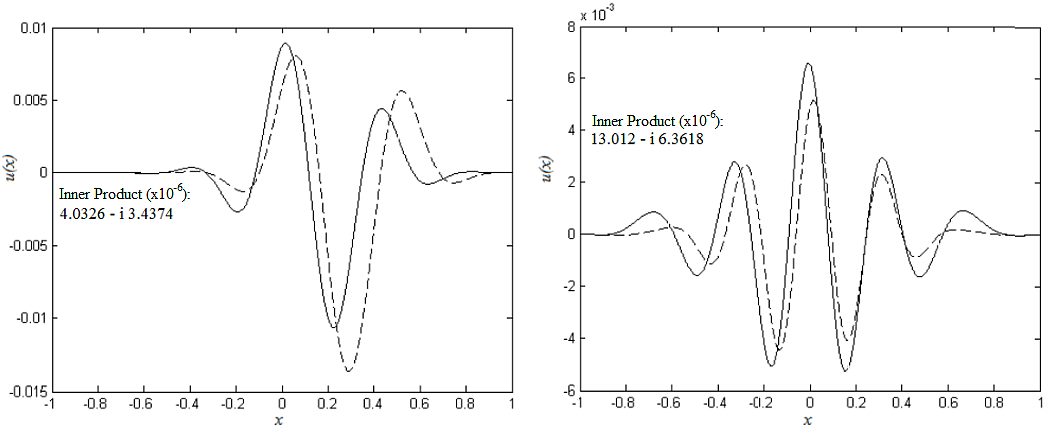}}
	\caption{Two typical sets of nonnormal eigenmodes in the solution of eq. (\ref{Qsol}) for $q=1.5$ (Keplerian flow).
Solid and dashed lines represent the real parts of two different eigenvectors.
Other parameters are $k_y=0.4$, $k_z=1.6$, $Re=2000$ and $B_p\equiv (0,0.3,0.3)$.
}
	\label{nonorvec}
\end{figure}

We assume the form of perturbations to be
\begin{equation} \label{pert}
	f(x,y,z,t)\rightarrow f(x,t)e^{i (k_y y+k_z z)},
\end{equation}
where $f(x,y,z,t)\equiv u,\zeta,B_x,\zeta_B$.
On substituting the form for various perturbation fields from eq. (\ref{pert}) to eqs. (\ref{os}), (\ref{sq}),
(\ref{mos}) and (\ref{msq}), we can write the resulting 
equations in the form of an eigenvalue problem such that
\begin{equation} \label{eigenval}
\frac{\partial Q}{\partial t}=- i \hat {\cal M} Q, \quad 
	Q=\begin{pmatrix}
		u(x,t)\\ (1/\sqrt{4\pi})B_x(x,t)\\ \zeta(x,t)\\ (1/\sqrt{4\pi})\zeta_B(x,t)
	\end{pmatrix}, \quad
	\hat {\cal M}=\begin{pmatrix}
		M_{1} & M_{2} & M_{3} & 0\\
		M_{4} & M_{5} & 0 & 0\\
		M_{6} & 0 & M_{1} & M_{2}\\
		0 & M_{7} & M_{4} & M_{5}\\
	\end{pmatrix},
\end{equation}
$
\hfill
\begin{array}{lcllcllcllcl}
		M_{1} & = & {-\left({\frac{(\partial_x^2 - k^2)}{i Re}+xk_y}\right)}, &
		M_{2} & = & \frac{-(\overrightarrow{B_p}.\overrightarrow{k})}{4 \pi}, &
		M_{3} & = & \frac{1}{\partial_x^2 - k^2}{\left(\frac{2 k_z}{q}\right)}, & \, & \, & \, \\
		M_{4} & = & {-(\overrightarrow{B_p}.\overrightarrow{k})},  &
		M_{5} & = & {-\left({\frac{(\partial_x^2 - k^2)}{i Rm}+xk_y}\right)}, &
		M_{6} & = & \left({1-\frac{2}{q}}k_z\right), &
		M_{7} & = & -k_z,
\end{array}
\hfill
$
\\
where $\overrightarrow{k}=(0,k_y,k_z)$. 
On further considering
\begin{equation} \label{Qsol}
	Q(x,t)=\sum_{j=1}^{\infty} C_j e^{-i\sigma_j t}\tilde{Q}(x),
\end{equation}
$\hat {\cal M}$ follows the eigenvalue equation $\hat {\cal M}Q_j=\sigma_j Q_j$,
where $\sigma_j$ is of the form $\sigma_j=\sigma_{Rj}+i \sigma_{Ij}$.
Note importantly that although the form of the solution in eq. (\ref{Qsol})
is chosen in the spirit of normal-mode expansion, $\hat {\cal M}$ is not 
self-adjoint and, hence, the resulting set of eigenmodes ($Q_j$-s) may be highly 
sensitive to the choice of perturbations and the eigenfunctions may be nearly linearly dependent
(see, e.g., \cite{schmidhenningson}), thus effectively called nonnormal.
Figure \ref{nonorvec} represents the real parts of two pairs of stable eigenvectors,
for a Keplerian flow as an example, whose inner-product is non-negligible.
It clearly shows the nonnormal nature of eigenmodes.
The shapes of eigenmodes indicate how nonnormal they are, when for a self-adjoint operator
the eigenmodes are perfectly normal (orthogonal) with zero inner-product.
Note that the inner-product is computed for the complete nonnormal eigenvectors (and not just their real
parts) in each pair.
 
\begin{figure}[h]
	\centering
	\fbox{\includegraphics[width=0.5\linewidth]{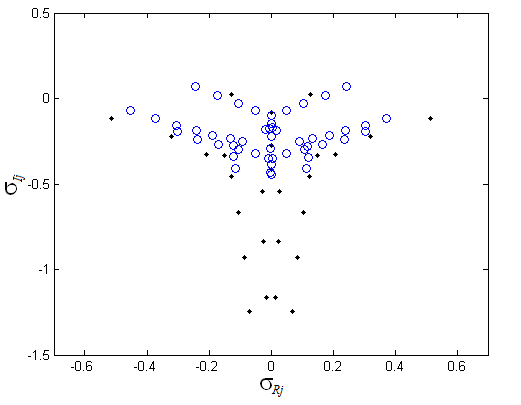}}
	\caption{Eigenspectra for $Re=100$ (black points) and $1000$ (blue 
circles) for $q=1.5$ (Keplerian flow).
Other parameters are $k_y=0.4$, $k_z=1.2$ and $B_p\equiv (0,0.3,0.3)$.
}
	\label{eigcom}
\end{figure}

Now, these eigenvalue equations are well studied for purely hydrodynamical cases (Orr-Sommerfeld and Squire equations)
without and with rotation \cite{man2005}, and have been shown to have no unstable modes for the velocity and vorticity
fields. However, the above set of four equations, in fact, has an unstable set of solutions corresponding to MRI, which is explored in depth in 
subsequent sections.

\subsection{Transient Energy Growth} \label{transen}

The application of TEG was initially explored by Farrell \cite{farrell88}, Reddy \& Henningson \cite{reddyhenningson1993}, 
Trefethen and his collaborators \cite{sc} to explain observed instabilities in linearly stable Couette flow
and Poiseuille flow (see also \cite{berk}). Later on, the concept was further modified and applied to the cases of rotating shear flows
\cite{man2005,chag,umur,yecko}. Here we further extend and explore this in the presence of magnetic fields. Such an exploration
was pursued recently in the Lagrangian formulation \cite{nath2015}. In the Eulerian formulation, here we will be  
in a position to explore various kinds of eigenspectra, depending on, e.g., the values of $q$, $\vec{k}$ and $\vec{B_p}$,
and their role in controlling TEG of perturbation. Note that the structure of eigenspectra is also related to 
the nature of nonnormality, which further controls the perturbation energy growth factor, and emergence of 
nonlinearity and plausible turbulence in the flows.
The expression for perturbation energy growth is given by
\begin{equation}
	G=\frac{1}{2V}\iiint_V \left((u^2 + v^2 + w^2) + \frac{1}{4 \pi}({B_x}^2+{B_y}^2+{B_z}^2)\right) \,dx\,dy\,dz,
\end{equation}
where $V$ is the volume of the chosen system (e.g. shearing box). 
Using the solution form given by eq. (\ref{Qsol}) in eq. (\ref{eigenval}), we can write
\begin{equation}
	Q(x,t)=e^{-i{\cal L} t}Q(x,0)
\end{equation}
and, hence, maximum growth in perturbed energy is expressed as
\begin{equation}
	G_{max}=\max {\left[\frac{{\|{Q(x,t)}\|_2}^2}{{\|{Q(x,0)}\|_2}^2}\right]}={\|e^{-i{\cal L}t}\|_2}^2,
\end{equation}
where $\|\ldots\|_2$ refers to the 2-norm/Euclidean norm. The 2-norm can be numerically computed via a scheme involving optimization of the coefficients $C_j$, as outlined in previous work \cite{man2005,reddyhenningson1993}. Briefly put, 
the perturbation energy can be written as a sum of complex conjugate products of the 4 perturbation variables and derivative
of two of the variables
($v,w,B_y,B_z$ can be substituted in terms of $\zeta, \zeta_B, \partial u/\partial x, \partial B_x/\partial x$), 
which can be arrived at by multiplying $\hat{Q}(x)$ with its complex conjugate resulting in a hermitian matrix 
$\hat{Q}_{ij}$. Further, to obtain the expression for optimum growth, $\hat{Q}_{ij}$ is decomposed in terms of 
a lower-triangular matrix $W$ by means of a Cholesky-decomposition, which is then used to write the final 
expression of energy growth for $K$ eigenvalues as
\begin{equation}
G_K(t)={\|We^{-\Sigma_Kt}W^{-1}\|_2}^2,
\end{equation}
where $\Sigma_K$ is the diagonal matrix with elements as $K$ eigenvalues.

\subsection{Numerical Considerations} \label{numcond}

We use the publicly available Chebfun-MATLAB package \cite{chebfun} 
(with the appropriate modification for the present purpose) 
to perform the numerical computations for the eigenvalue system 
described in the previous section. Exploration of visco-resistive MHD stability 
using Chebyshev polynomials is nothing new however (see, e.g., \cite{pnas}).
Using the example given on-line, based on Orr-Sommerfeld operator,
we form the eigensystem for the magnetized version of coupled 
Orr-Sommerfeld and Squire equations including the effect of rotation, for
the present computation purpose.
Note that beyond a certain number of eigenvalues and, hence, 
the matrix dimension ($60-80$, depending on the value of $q$), the Cholesky-decomposition of $\tilde{Q}(x)$ 
for the purposes of calculating the 2-norm cannot be done because the determinant of the 
matrix $\hat{Q}_{ij}$ does not remain positive definite 
on using the Chebfun package, which poses problem to compute TEG. 
However, once such a situation arises, the code truncates 
the number of eigenvalues and, hence, the matrix dimension to assure the determinant of 
$\hat{Q}_{ij}$ to be positive.
The last few eigenvalues, in our analyses, however, do not seem to make any practical difference in 
computing the value of TEG. Although Chebfun software is probably 
not best suited for accuracy purposes, for the present purpose this does not pose 
much hindrance where the aim is to qualitatively understand the effects of nonnormality 
in shearing MHD flows (see \cite{mhdbook2} to understand the requirement of 
better linear algebra software which could handle up to thousands of modes).


\section{Simpler Analytical Exploration} \label{anal}

Before we discuss the numerical results in detail, here we try to understand some fundamental 
properties of perturbation based on simpler approximate analytical solutions. 
Note that due to the assumption of incompressibility, the fast magneto-acoustic modes (and its
modifications) have already been eliminated and we are left with the slow modes which are 
degenerate with the Alfv\'{e}n modes. 
See \cite{mhdbook1} for the detailed description of magneto-acoustic modes. In brief, the 
magneto-acoustic modes are generated in MHD flows under perturbation, which are four in number
including forward going and backward going modes. 

For ease of understanding, let us consider the simpler plane wave perturbations, of form 
$Q=\exp[i(k_x x-\sigma t)]$,
unlike the more generalized choice as given in eq. (\ref{Qsol}), which indeed will be used 
in subsequent sections.
The corresponding dispersion relation can be obtained by substituting the above plane wave perturbation 
into eq. (\ref{eigenval}), taking its determinant such that
$|\hat{\cal M}-\sigma I|=0$ (where $I$ being the unit matrix), given by
\begin{equation} \label{spectral}
(M_{1}-\sigma)^4 - (M_{2}M_{4} + M_{3}M_{6})(M_{1}-\sigma)^2 - M_{2}M_{3}M_{4}M_{7} = 0,
\end{equation}
where, without loss of any crucial physics for the present purpose, we assume $Re = Rm$, 
giving us $M_{1}=M_{5}$. 

The choice of plane wave perturbation allows us to substitute
$\partial_{x}$ with $ik_x$. We now define Alfv\'{e}n frequency 
$\omega_A=(\overrightarrow{v_A}.\overrightarrow{k})$, where Alfv\'{e}n velocity $v_A=B_p/\sqrt{4\pi}$
(as per our choice of dimension, mentioned in \S II.A),
and $k_x^2 + k_y^2 + k_z^2 = \tilde{k}^2$, and eq. (\ref{spectral}) reduces to
\begin{equation} \label{planespectral}
\left(\sigma + \frac{\tilde{k}^2}{i Re} + x k_y\right)^2 = \left(\frac{\omega_{A}^2}{2} - 
\frac{k_z (1 + k_z)}{q \tilde{k}^2}\right) 
\pm \sqrt{\left(\frac{\omega_{A}^2}{2}\right)^2 + \left(\frac{k_z (1 + k_z)}{q \tilde{k}^2}\right)^2 
+ \frac{\omega_A^2 k_z (k_z
- 1)}{q \tilde{k}^2}}.
\end{equation}

From the above equation, we easily see that on taking the axisymmetric ($k_y = 0$) ideal MHD 
($Re \to \infty$) for plane Couette ($q \to \infty$) flow, we recover the Alfv\'{e}n modes,
which manifest in the eigenspectra as symmetric modes on the real axis in complex plane (see 
the eigenspectra in subsequent sections), provided magnetic field is not 
insignificant. As $\sigma$ here is chosen real, all the modes are stable, which is 
indeed the case for plane Couette flow. 
The inclusion of rotation (finite $q$) gives rise to 
two additional sets of solutions, which can be interpreted as modified
Alfv\'{e}n modes that results in stable and unstable modes, where part of them are overlapping along the 
$y$-axis (see, e.g., top-left panel of Fig. \ref{kz15} below). If we consider visco-resistive effects 
(finite $Re$ and $Rm$), we see that the spectra shift down in the
complex plane because of an additional negative imaginary term $\frac{k^2}{iRe}$, which decreases 
in magnitude on increasing $Re$: see Fig. \ref{eigcom}. Finally, considering the non-axisymmetric case 
($k_y \neq 0$) causes the inclusion of the coordinate dependent shear term arising due to background flow 
(nonzero $x k_y$), which causes forward and backward Doppler shifting of the present modes 
\cite{mhdbook2}. This shows up in the eigenspectra as splitting of the vertical branches (see, e.g.,
Fig. \ref{ky15} below), 
a very characteristic feature of rotating shear flows in general (see also \cite{pp}).
Note also that as $k_z\rightarrow 0$, the various branches tend to overlap 
due to decreasing rotational effect and modes tend to become
degenerate, as is clear from eq. (\ref{planespectral}). See the evolution of eigenspectra with the 
change in $k_z$ shown below.

\section{Numerical Results} \label{numrel}

We consider the following cases:
\begin{itemize}
	\item Keplerian disk, i.e., $q=1.5$,
	\item constant angular momentum flow, i.e., $q=2$,
	\item plane Couette flow, i.e., $q\to \infty$.
\end{itemize}

\subsection{Unstable Solutions}

Chandrasekhar \cite{book} already explored the regimes of instability in MHD Taylor-Couette flow for 
a variety of cases which follows a slightly different formulation than ours. We consider the general 
non-axisymmetric case and, on fixing the values of $Re$ and $Rm$ (with $Re=Rm$ for ease of analysis and 
also because their magnitudes, in main applications under consideration, are roughly the same), obtain an approximate 
regime of instability in the $(k_y,k_z,B_p)$ parameter space. We further choose,
$B_{p1}=0$ (for satisfying the original 
unperturbed equation) and $B_{p2}=B_{p3}$ (because that does not pose hindrance on any new physics).
The unstable solutions, however, do not exist for plane Couette flow discussed below.
\begin{figure}[h]
	\centering
	\fbox{\includegraphics[width=\linewidth]{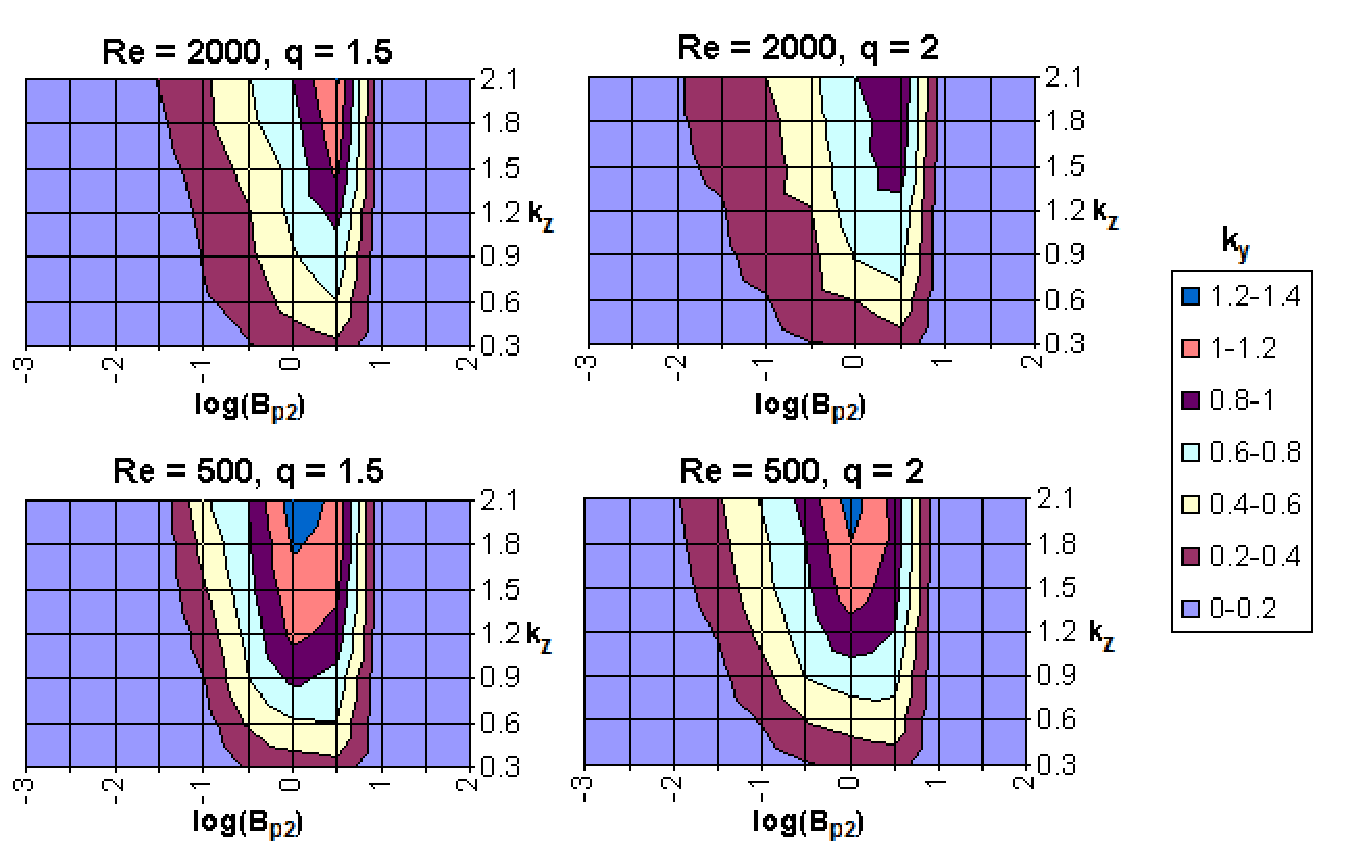}}
	\caption{Contours of constant $k_y$, demarcating the instability (MRI)
region, in the $B_{p2}-k_z$ plane for two values of 
$Re$ and $q$. The contours are for $k_y=0.2,0.4,0.6 \ldots$, moving from outermost to innermost region.}
	\label{fig1}
\end{figure}

The various contours in Fig. \ref{fig1} show the regions of instability as functions of $k_y, k_z$ and $B_{p2}$. 
At very low $B_{p2}$, there are practically no unstable flows. However, at e.g $q=1.5$, for $B_{p2}\sim 0.03$, at 
a low value of $k_y$, the flow starts to exhibit unstable behaviour due to MRI. The value of $k_y$ leading 
to instability continues to increase with increasing $B_{p2}$ till a critical value of $B_{p2}$, above which, again, 
there is no instability. 
This feature is consistent across both the cases of $Re$ considered and with the two different types of flow considered 
as well. Moreover, above a certain value of $k_z$ (e.g. $k_z \sim 40$  for $k_y=0$ and $B_{p2}=0.3$), the unstable 
region vanishes again.

The order of magnitude of most of the unstable (positive $\sigma_{Ij}$) eigenvalues ranges as 
$0\lesssim \sigma_{Ij}\lesssim 1$. 
Hence, on comparing this range with the values of $\omega_A$, 
obtained from the range of $k_y, k_z$ and $B_{p2}$
giving rise to instability, one finds that their orders of magnitude match and, hence, they 
correspond to MRI (indeed the best MRI growth rate corresponds to 
$\sigma = \overrightarrow{k}.\overrightarrow{v_A}$ \cite{rmp}).
Note that the contours are only rough boundaries and, hence, their jagged appearance. 
Thus, they illustrate the behaviors exhibited approximately.

\subsection{Transition to Stable Solutions with Transient Energy Growth}

In the case of stable solutions surrounding the unstable zone shown in Fig. \ref{fig1}, certain trends are 
noticed in TEG as well as the eigenspectra. For evaluating the variance of energy growth with changing 
wavenumber from its value corresponding to unstable region, for given values of $Re~(=2000)$ and 
$B_p~(\equiv 0,0.3,0.3)$, we fix one of the components of wavenumber and vary the other component. 
The following cases were considered:
\begin{enumerate}
\item fixed $k_y \, (=0.4)$ and $k_z$ varied as $1.6, 1.2, 0.8, 0.4, 0$,
\item fixed $k_z \, (=1.2)$ and $k_y$ varied as $0, 0.4, 0.8, 1.2, 1.6$.
\end{enumerate}
For these cases, eigenspectra as well as energy growths are shown to 
reveal their evolution. 

\subsubsection{Keplerian disk ($q=1.5$)}
\begin{figure}[!ht]
	\centering
		\fbox{\includegraphics[width=\linewidth]{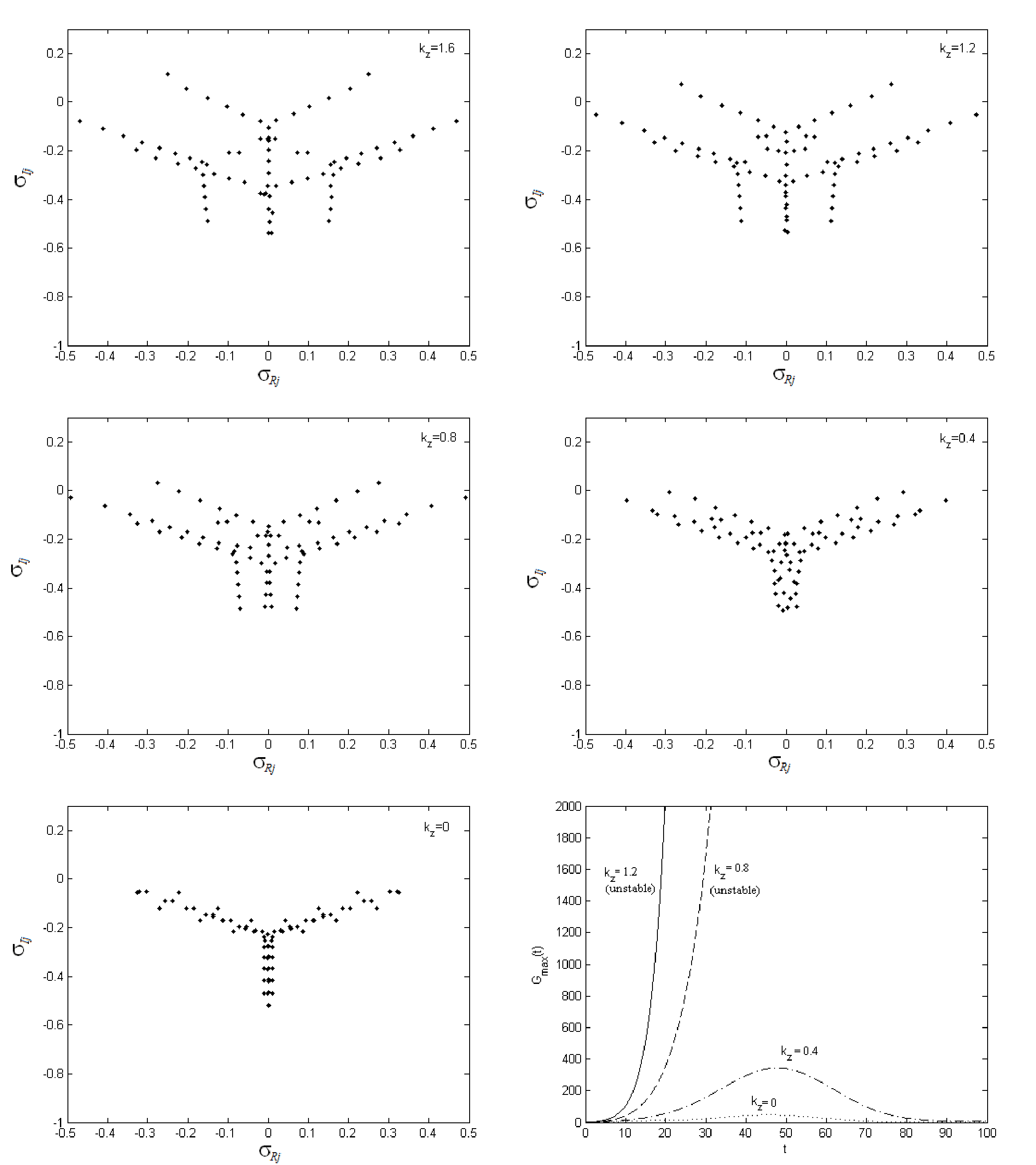}}
		\caption{Eigenspectra and energy growth for $q=1.5$ with a fixed $k_y=0.4$, when $Re=2000$
and $B_p\equiv(0,0.3,0.3)$.}
		\label{ky15}
\end{figure}
Figure \ref{ky15} shows the eigenspectra in the complex plane, with the vertical axis corresponding to the nature 
of the eigenmodes (if above $0$, the modes are unstable and vice versa) and 
horizontal axis corresponding to the wave 
part of the solution, as per the description in eq. (\ref{Qsol}).
Considering the fixed $k_y$ case first, one can see in Fig. \ref{ky15} that
on moving away from the highly unstable region (by decreasing $k_z$ in the
contour plots for fixed $k_y$ and $B_{p2}$, shown in Fig. \ref{fig1}) towards the stable region, 
the eigenspectra start to become degenerate, as discussed in \S III, and also tend to become fast-decaying.
Consequently, one can see from the growth curves that
the peak TEG decreases regularly, from a maximum of about $\sim 400$ for $k_z=0.4$. 
Note that the amplitude of TEG is directly correlated with the number of slowly-decaying low-frequency 
modes, which may allow an optimal linear combination over sufficient timescales to exhibit substantial TEG.

\begin{figure}[h]
	\centering
		\fbox{\includegraphics[width=0.95\linewidth]{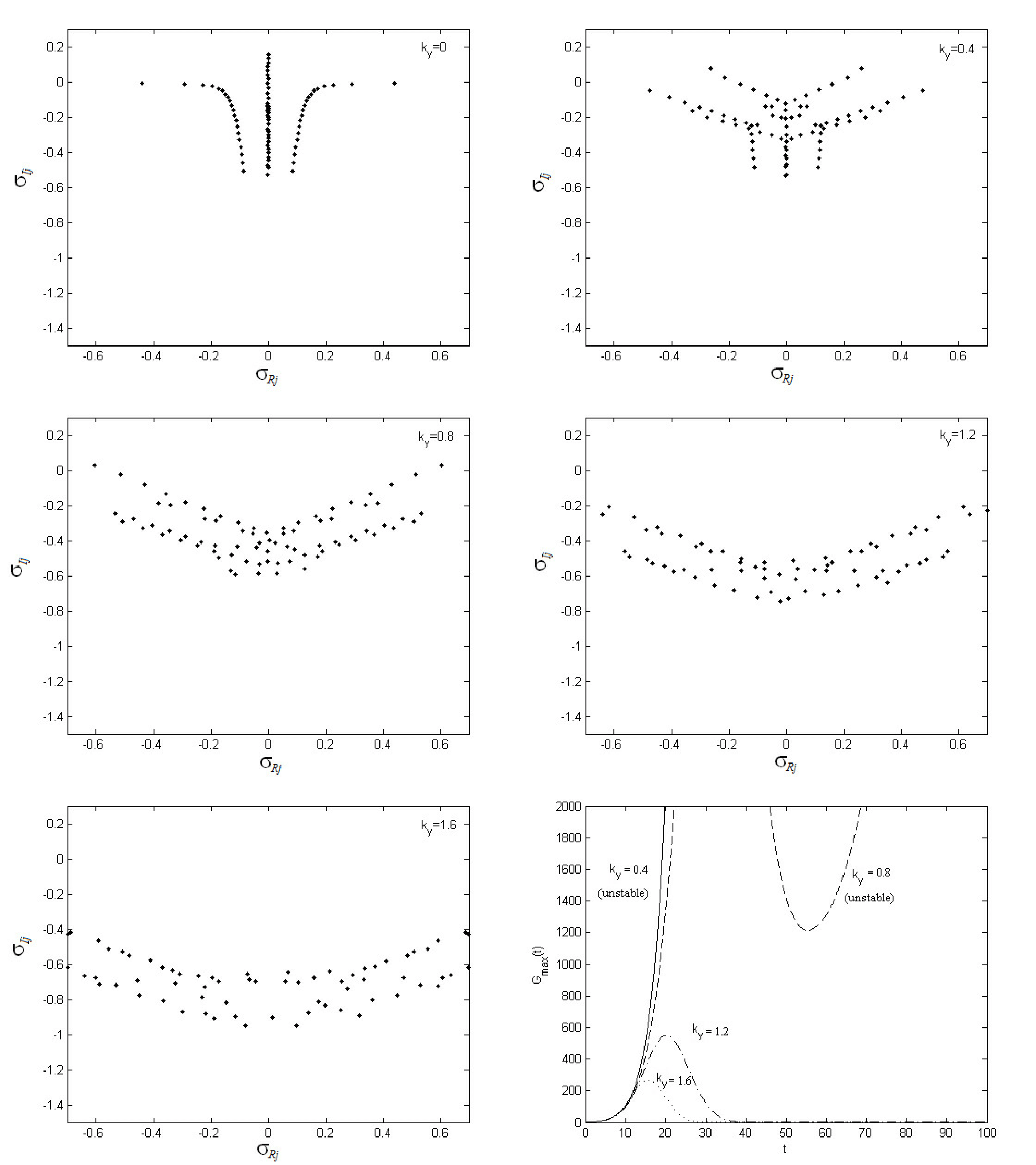}}
		\caption{Eigenspectra and energy growth for $q=1.5$ with a fixed $k_z=1.2$, when $Re=2000$
and $B_p\equiv(0,0.3,0.3)$.}
		\label{kz15}
\end{figure}


In the case of fixed $k_z$ (Fig. \ref{kz15}), the eigenspectra 
for $k_y=0$ do not have Doppler-shifted modes. On moving $k_y$ away from unstable region    
(by increasing $k_y$ in the direction away from the plane of the contours,
shown in Fig. \ref{fig1}), the spectra split to give 
the characteristic
Y-shape observed for all the four separate branches, as per the discussion in \S III. 
Continual increase of $k_y$ results in further
shifting of spectra towards the sides (because of the $x$-coordinate dependent velocity term in eq.
(\ref{planespectral})). 
As a result (as seen in bottom-right panel), the peak TEG decreases regularly 
(maximum for a stable system being $\sim 600$), although the rate of trailing of the TEG curve 
is much higher than in the case of fixed $k_y$. An interesting feature is seen for
$k_y=0.8$, which exhibits an initial peak then a minimum, followed by exponential growth. 
Such a situation arises when the growth rate of an unstable mode is lower than the initial TEG rate,
as discussed earlier \cite{nath2015} in Lagrangian formulation and also seen in the linear perturbation 
of plane Poiseuille flow \cite{reddyhenningson1993} at, e.g., $Re=8000$, $k_y=1$ and $k_z=0$.

\begin{figure}[h]
	\centering
		\fbox{\includegraphics[width=0.50\linewidth]{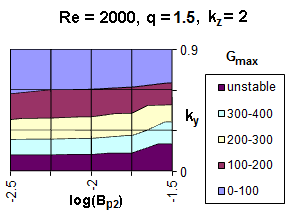}}
		\caption{Maximum TEG contours with $k_z=2$ for Keplerian flows. The contours from 
top to bottom regions are moving from the regions with small to large TEGs and finally to unstable region.}
		\label{cont15}
\end{figure}

Contours of Fig. \ref{cont15} show how the maximum TEG increases with the change of $k_y$ and finally
leading to linear instability below certain $k_y$. 
For the chosen range of magnetic field, maximum TEG turns out to be smaller, similar to the 
nonmagnetic cases reported earlier \cite{man2005}.

\subsubsection{Taylor-Couette flow ($q=2$)}
In the case of Taylor-Couette or constant angular momentum flow, we observe a similar trend in the eigenspectra and
TEGs as compared with Keplerian disk, except for one major difference, i.e. the actual growths are much higher.
\begin{figure}[h]
	\centering
		\fbox{\includegraphics[width=0.95\linewidth]{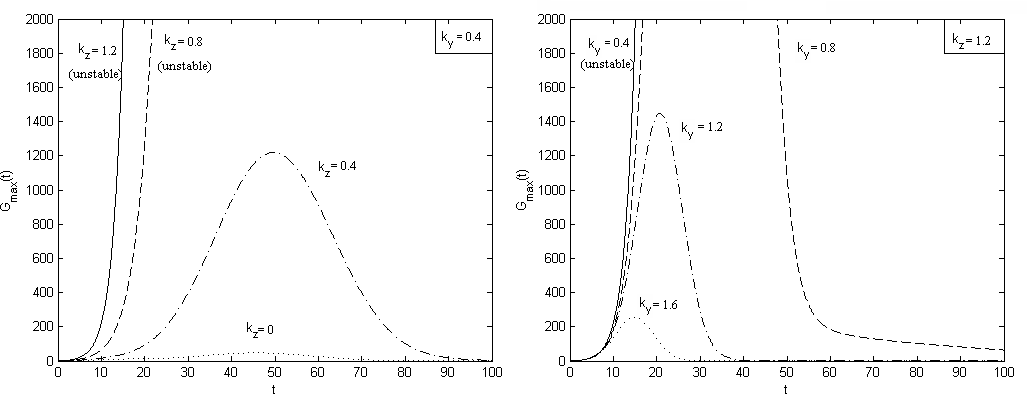}}
		\caption{Energy growth for $q=2$ with a fixed $k_y=0.4$ and a
fixed $k_z=1.2$, when $Re=2000$
and $B_p\equiv(0,0.3,0.3)$.}
		\label{kykz2}
\end{figure}
The fixed $k_y$ case, 
as seen in Fig. \ref{kykz2}, 
the maximum energy growth 
is much higher than the corresponding growth for $q=1.5$ case ($\sim 400$). This trend is comparable to earlier result
\cite{man2005} for hydrodynamic case comparing energy growth in these two different flows.
In the case of fixed $k_z$, 
the difference in the level of peak energy growth is apparent from Fig. \ref{kykz2}, 
exhibiting peak growth $\sim 1400$ (for the stable system with $k_y=1.2$) compared to $\sim 500$ for $q=1.5$.
The main reason for this difference in peak TEG between $q=1.5$ and $q=2$ cases is that, in the latter case, 
the second term in eq. (\ref{sq}) vanishes. This term, arising from the Coriolis effect, otherwise has a diminishing 
effect on the field perturbation due to generation of epicyclic fluctuations in the flow.

\begin{figure}[h]
        \centering
                \fbox{\includegraphics[width=0.5\linewidth]{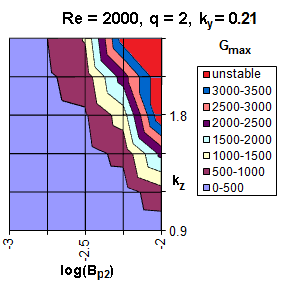}}
                \caption{Maximum TEG contours with $k_y=0.21$ for $q=2$ cases.
The contours from
left to right regions are moving from the regions with small to large TEGs and finally to unstable region.
}
                \label{cont2}
\end{figure}

Contours of Fig. \ref{cont2} show how the maximum TEG increases with the change of $k_z$, 
for the chosen range of magnetic field. 
We choose $k_y$ in such a way that TEG appears to be maximum. 
Interestingly, with the increase of $k_z$, first maximum TEG increases and then the flow
becomes unstable.

\subsubsection{Plane Couette flow ($q \to \infty$)}
Plane Couette flow does not show any unstable mode, even in the presence of magnetic field. 
Indeed it is known that in order to have MRI, the flow must exhibit rotation and magnetic field both together.
There is, however, considerable TEG, with peak $\gtrsim 10^4$ even at $Re=2000$, in the presence of 
magnetic field, which reveals nonlinearity and plausible turbulence.
\begin{figure}[h]
	\centering
	\begin{minipage}{0.5\linewidth}
		\includegraphics[width=\linewidth]{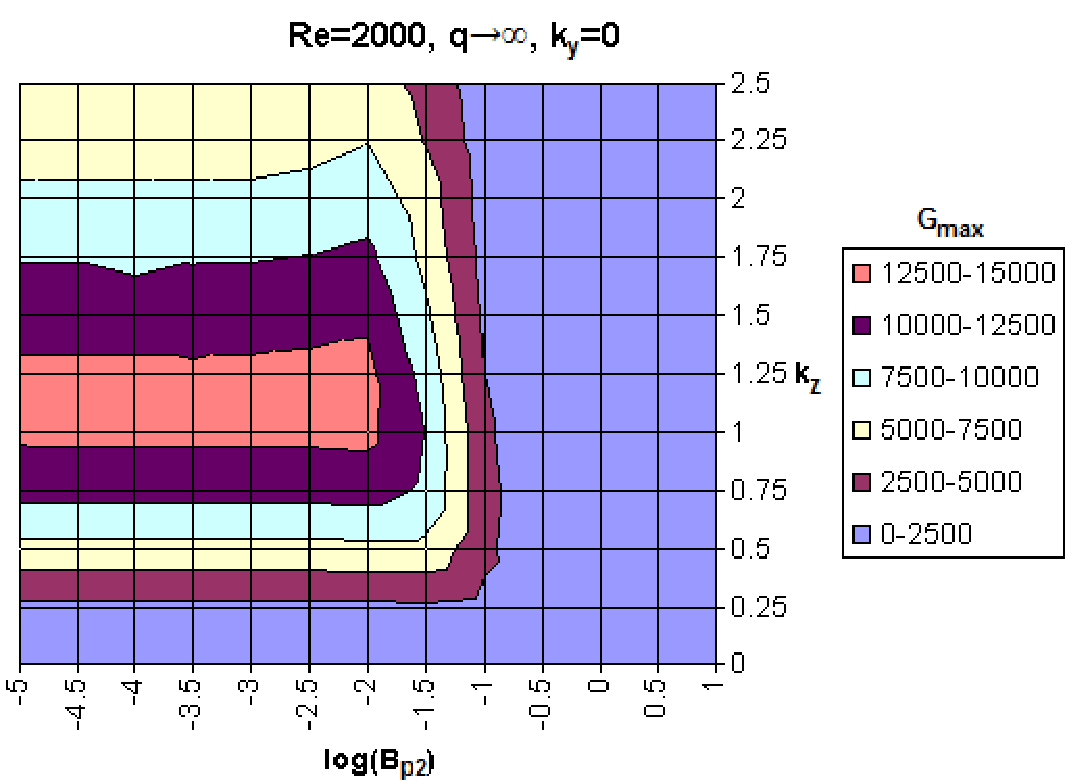}
		\caption{Maximum TEG contours with $k_y=0$ for plane
Couette flow.}
		\label{qinfcont}
	\end{minipage}
	\begin{minipage}{0.85\linewidth}
		\includegraphics[width=\linewidth]{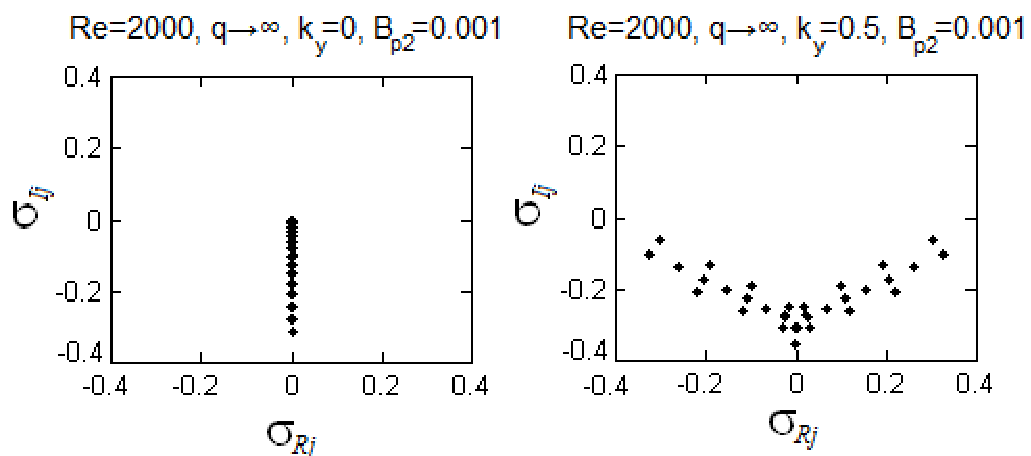}
		\caption{Variation of eigenspectra with $k_y$ for 
plane Couette flow, where $k_z=1.2$.}
		\label{qinfeigs}
	\end{minipage}
\end{figure}

Figure \ref{qinfcont} shows the peak values of TEG. We choose $k_y=0$, because the maximum energy growth 
is revealed around this $k_y$. 
To the right hand side of $B_{p2} \sim 1$, the peak of TEGs drastically decreases ($< 20$).
Figure \ref{qinfeigs} shows that on increasing $k_y$ from $0$ onwards, when the
magnetic field is weaker ($\omega_A$ is smaller), the eigenspectrum splits from a shape of single 
vertical branch to Y-shaped spectra due to emergence of Doppler shifted modes.
This also results in lowered energy growth as a consequence of 
deviation from nonnormality.

\subsubsection{High magnetic field regime}
\begin{figure}[h]
	\centering
		\fbox{\includegraphics[width=0.95\linewidth]{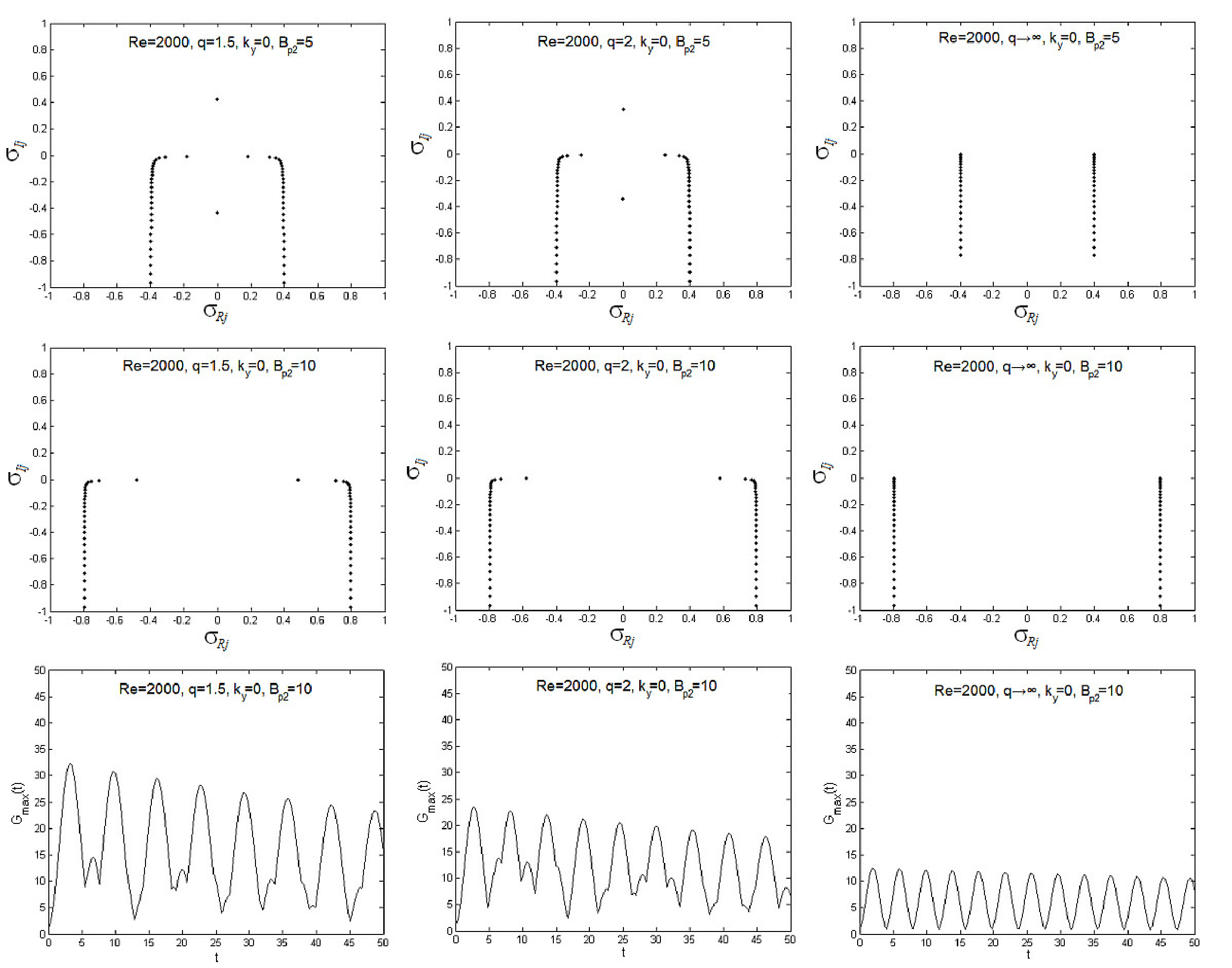}}
		\caption{Eigenspectra in the presence of higher magnetic fields and corresponding energy growth,
for $q=1.5$, $2$ and $\infty$ (plane Couette flow) with $k_y=0$.}
		\label{largeB}
\end{figure}
In the range of wavenumbers $k_y,k_z \sim 0-3$, the magnetic field $B_{p2}$ higher than $\sim 10$ corresponds to 
stable flows. Interestingly, in this parameter region, even TEG is extremely reduced. Even in the case of 
plane Couette flow ($q\rightarrow\infty$), for $B_{p2} \gtrsim 10$, Fig. \ref{qinfcont} shows practically no growth. 
The resulting solutions are of damped oscillatory type. We can try to get some insight into why this could happen 
by observing the typical eigenspectra shown in Fig. \ref{largeB}. The various intermingling branches in all previous 
eigenspectra exhibited substantial TEG have here separated into two distinct and widely spaced branches, which 
further reduce nonnormality substantially. Apart from some minor differences, this separation of branches 
is consistent with all three types of flows.
Stability in the presence of high magnetic fields can also be understood due to
the emergence of high frequency, rapidly oscillating modes.
While one can always increase the background magnetic field to increase the value of $\omega_A$,
which can be inferred from the first parentheses
at right hand side in eq. (\ref{planespectral}), the magnitude of the second term of the 
parentheses (the one involving $q$ and $k_z$)
is limited to $1+3/[2(2+\alpha)]$, where $\alpha^2=1+1/(k_x^2 + k_y^2)$. Hence, high field stabilizes 
the system and kills of TEG by giving
rise to rapidly oscillating modes, regardless of the wavenumbers.\\ \\

\begin{figure}[h]
	\centering
		\fbox{\includegraphics[width=0.95\linewidth]{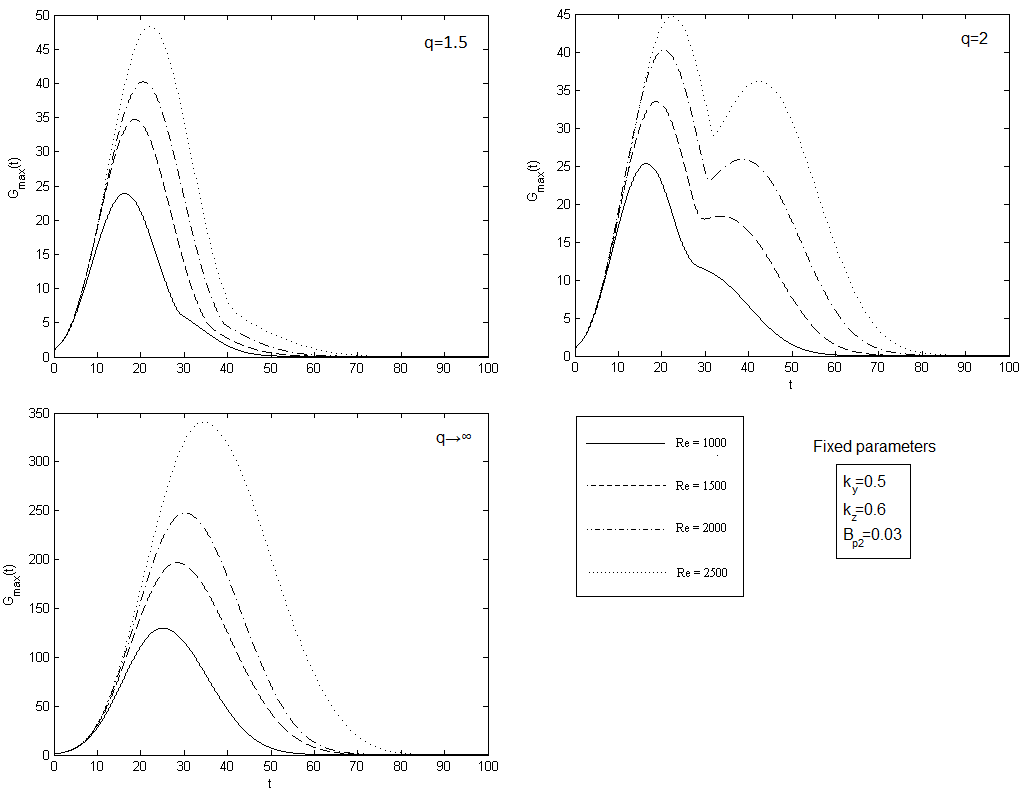}}
		\caption{Variation of TEG peak with $Re$.}
		\label{revar}
\end{figure}

Most of the results presented above are for a fixed $Re$. With the increase of $Re$, keeping 
other parameters intact, however, the peak of TEG increases, which has been demonstrated in 
Fig. \ref{revar} for all three typical flows considered here. 
As $Re$ in accretion disks is huge ($\gtrsim 10^{15}$ \cite{plb}), it is expected to exhibit
huge TEG with very large peak growths to produce nonlinearity and subsequent turbulence.
The second peak for $q=2$ cases 
is due to the choice of same perturbation for all values of $q$. Note that when a constant angular 
momentum flow exhibits best growth in the presence of vertical perturbation, the Keplerian 
flow needs a two-dimensional perturbation to reveal best growth. As a result, due to several 
competing modes, in particular at higher values of $Re$, $q=2$ cases produce two peaks. 

%

\section{Summary and Conclusion}\label{summa}

We have explored and compared how linear instability and TEG may arise in MHD flows for
Keplerian disk, constant angular momentum flow and plane Couette flow in terms of an eigenvalue
formulation of the shearing box model.
The system considered is the incompressible visco-resistive MHD flow following the
Orr-Sommerfeld and Squire operator formulation, supplemented by the Coriolis
effects and magnetic fields.
In terms of spectral decomposition, such a system, by design, does not exhibit any fast
magneto-acoustic modes and the underlying slow modes are degenerate with the Alfv\'{e}n modes which,
in the presence of rotation, may also exhibit MRI. The basic trends in the system can be understood by a simple 
plane wave perturbation analysis. The incorporation of visco-resistive effects 
(by having a finite $Re$ and/or $Rm$) and non-axisymmetry (non-zero azimuthal 
wavenumber $k_y$) result in a variety of modes (modifications of the basic Alfv\'{e}n 
modes), which manifest physically as lesser number of unstable modes as well as lowered TEG.

It seems that, in the case of stable systems, the amplitude of TEG is directly correlated 
with the number of slowly-decaying low-frequency modes, since these are the modes that may 
allow an optimal linear combination over sufficient timescales to exhibit substantial TEG.
Perturbations with non-axisymmetric component (non-zero $k_y$) tend to get sheared by 
the flow, resulting in high-frequency Doppler-shifted oscillatory modes. These modes have a 
lower possibility of optimal linear combinations and, hence, do not show significant TEG. 
Since the flows under consideration have plane (and linear) shear, axisymmetric perturbations 
therein remain unaffected and exhibit instability or substantial TEG. 
We posit that high frequency oscillations do not allow the
modes to have optimal linear combination over a given time range, which is necessary for large TEG.

Perturbations with vertical component are 
affected by both rotation (leading to non-zero vorticity; essential for turbulence) and the 
background magnetic field strength. A certain finite range of these perturbations allows for 
instability and significant TEG.
This range (which is also dependent on                               
the value of background magnetic field) specifies where MRI and where significant TEG can occur.
Beyond this range (above certain value of vertical wavenumber $k_z$ for a given magnetic field), 
the magnetic field stabilizes the flow and in the absence of vertical perturbation the flow becomes irrotational. 
In either of the cases, instability is reduced (having less number of unstable modes with
lower growth rates) as well as TEG is decreased.

Last, strong background magnetic fields tend to 
have a stabilizing effect on the perturbations, which can be understood by invoking the 
``rod''-like nature of these fields, compared to the ``spring"-like nature of 
weak fields governing MRI. What is more interesting to note is that 
these strong fields also kill off TEG. 

The type of modes that we consider is limited by the assumption of incompressibility, the shearing
box model (which ignores the effects of curvature) and plane wave perturbation in the azimuthal
and vertical directions. A lot of work, which includes some of these consideration, but limited to the
scope of linear stability analysis, is already present in literature \cite{jpp,apjl}. A more complete
picture of MHD TEG may emerge with the study of compressible flows in cylindrical coordinates
with more generalised perturbations. \\ \\

T.S.B. would like to thank the Department of Physics, Indian
Institute of Science, Bangalore, for providing support to pursue his Master's thesis, where
this research was conducted. A partial financial support from the project 
with research Grant No. ISTC/PPH/BMP/0362 is acknowledged. 
Finally thanks are due to the referees for their valuable critique and 
suggestions. 



\begin{thebibliography}{73}

\bibitem{pringle1981}
J.E. Pringle, ARA\&A \textbf{19,} 137 (1981)

\bibitem{shasun1973}
N.I. Shakura and R.A. Sunyaev, Astron. Astrophys. \textbf{24,} 337 (1973)

\bibitem{balbushawley1991}
S.A. Balbus and J.F. Hawley, Astrophys. J. \textbf{376,} 214 (1991)

\bibitem{velikhov1959}
E. Velikhov, J. Exp. Theor. Phys. \textbf{36,} 1398 (1959)

\bibitem{chandrasekhar1960}
S. Chandrasekhar, Proc. Nat. Acad. Sci. (USA) \textbf{46,} 253 (1960)

\bibitem{nath2015}
S.K. Nath and B. Mukhopadhyay, Phys. Rev. E \textbf{92,} 023005 (2015)

\bibitem{mhdbook1}
J.P. Goedbloed and S. Poedts, \textit{Principles of Magnetohydrodynamics} (Cambridge University Press, 2004)

\bibitem{fusion}
J.P. Freidberg, Rev. Mod. Phys. \textbf{54,} 801 (1982)

\bibitem{mhdbook2}
J.P. Goedbloed, R. Keppens and S. Poedts, \textit{Advanced Magnetohydrodynamics} (Cambridge University Press, 2010)

\bibitem{jpp}
C. Wang, J.W.S. Blokland, R. Keppens and J.P. Goedbloed, J. Plasma. Phys. \textbf{70,} 651 (2004)

\bibitem{pp}
J.P. Goedbloed, A.J.C. Beli\"en, B. van der Holst and R. Keppens, Phys. Plasmas \textbf{11,} 4332 (2004)

\bibitem{apjl}
R. Keppens, F. Casse and J.P. Goedbloed, Astrophys. J. \textbf{569,} L121 (2002)

\bibitem{ppastro}
J.P. Goedbloed, A.J.C. Beli\"en, B. van der Holst and R. Keppens, Phys. Plasmas \textbf{11,} 28 (2004)

\bibitem{sujitamit1}
S.K. Nath, B. Mukhopadhyay and A.K. Chattopadhyay, Phys. Rev. E \textbf{88,} 013010 (2013)

\bibitem{sujitamit2}
S.K. Nath and A.K. Chattopadhyay, Phys. Rev. E \textbf{90,} 063014 (2014)

\bibitem{orszag1971}
S.A. Orszag, J. Fluid Mech. \textbf{50,} 689 (1971)

\bibitem{trefethenembree}
L.N. Trefethen and M. Embree, \textit{Spectra and Pseudospectra} (Princeton University Press, 2005)

\bibitem{schmidhenningson}
P.J. Schmid and D.S. Henningson, \textit{Stability and Transition in Shear Flows} (Springer-Verlag New York, 2001)

\bibitem{orszag1972}
S.A. Orszag, Stud. Appl. Math. \textbf{51.3,} 253 (1972)

\bibitem{reddyhenningson1993}
S.C. Reddy and D.S. Henningson, J. Fluid Mech. \textbf{252,} 209 (1993)

\bibitem{man2005}
B. Mukhopadhyay, N. Afshordi, and R. Narayan, Astrophys. J. \textbf{629,} 383 (2005)

\bibitem{mc2013}
B. Mukhopadhyay and A.K. Chattopadhyay, J. Phys. A \textbf{46,} 035501 (2013)

\bibitem{farrell88}
B. Farrell, Phys. Fluids \textbf{31,} 209 (1988)

\bibitem{sc}
L. Trefethen, A. Trefethen, S. Reddy, and T. Driscoll, Science \textbf{261,} 57 (1993)

\bibitem{berk} C.D. Cantwell, D. Barkley and H.M. Blackburn, Phys. Flud. {\bf 22}, 034101 (2010)

\bibitem{chag}
A.G. Tevzadze, G.D. Chagelishvili, J.-P. Zahn, R.G. Chanishvili, and J.G. Lominadze, 
Astron. Astrophys. {\bf 407,} 779 (2003)

\bibitem{umur}
O.M. Umurhan, and O. Regev, Astron. Astrophys. {\bf 427,} 855 (2004)

\bibitem{yecko} P.A. Yecko, Astron. Astrophys. {\bf 425,} 385 (2004)

\bibitem{chebfun}
T.A. Driscoll, N. Hale, and L.N. Trefethen, \textit{Chebfun Guide} (Pafnuty Publications, Oxford, 2014);
L. Fox and I. B. Parker, \textit{Chebyshev polynomials in numerical analysis} (Oxford University Press, 1968)

\bibitem{pnas} R.B. Dahlburg, T.A. Zang, D. Montgomery, and M.Y.Hussaini,
Proc. Natl. Acad. Sc. {\bf 80,} 5798 (1983)

\bibitem{book}
S. Chandrasekhar, \textit{Hydrodynamic and Hydromagnetic Stability} (Dover, New York, 1961)

\bibitem{rmp}
S.A. Balbus and J.F. Hawley, Rev. Mod. Phys. \textbf{70,} 1 (1998)

\bibitem{plb}
B. Mukhopadhyay, Phys. Lett. B \textbf{721,} 151 (2013)



\end{thebibliography}
\end{document}